\documentclass{article}
\usepackage[utf8]{inputenc}
\usepackage[a4paper, total={6in, 9.5in}]{geometry}
\pdfoutput=1
\expandafter\let\csname equation*\endcsname\relax
\expandafter\let\csname endequation*\endcsname\relax
\usepackage{cite}
\usepackage{subfig}
\usepackage{subcaption}
\usepackage{amsmath}
\usepackage[protrusion=true,expansion=true]{microtype}
\usepackage{times}
\usepackage{graphicx,bm,xcolor,braket, tikz}
\usepackage[normalem]{ulem}
\usepackage{hyperref}
\usepackage{bookmark} % Add this line!
\hypersetup{colorlinks,linkcolor={blue},citecolor={blue},urlcolor={blue}}
\usepackage{amssymb}
\usepackage[utf8]{inputenc}
\usepackage[english]{babel}

\makeatletter
\def\graphicscale{\twocolumn@sw{0.3}{0.4}}
\def\graphicthreescale{\twocolumn@sw{0.3}{0.4}}

\newcommand{\be}{\begin{equation}}
\newcommand{\ee}{\end{equation}}
\newcommand{\bea}{\begin{align}}
\newcommand{\eea}{\end{align}}

\usepackage{amsmath}

\usepackage{dsfont}
\usepackage{mathrsfs}
\usepackage{fancyhdr}
\usepackage{esint}

\usepackage{comment}
\renewcommand{\subsectionmark}[1]{}
\begin{document}

\title{Double-weak-link interferometer of hard-core bosons in one dimension}
\author{Attila Tak\'acs$^{1,2}$, J\'er\^ome Dubail$^{3,2}$, Pasquale Calabrese$^{4}$}

\maketitle
\begin{center}
{\it
$^1$~Department of Physics, Faculty of Mathematics and Physics, University of Ljubljana, SI-1000 Ljubljana, Slovenia \\
$^2$~Universit\'e de Lorraine, CNRS, LPCT, F-54000 Nancy, France\\
$^3$~CESQ and ISIS (UMR 7006), University of Strasbourg and CNRS, 67000 Strasbourg, France\\
$^{4}$SISSA and INFN, Via Bonomea 265, 34136 Trieste, Italy\\
}
\vspace{10mm}
\end{center}
\begin{abstract}
We study the dynamics of a lattice hard-core boson gas released from a domain wall initial state in the presence of two weak links (defects). When the two defects are separated by a finite distance, the resulting density profile exhibits clear deviations from the standard Euler-scale hydrodynamic description of the gas, due to genuine quantum interference effects between the two defects. By analyzing the exact fermionic propagators, we show that repeated reflections at the defects give rise to interference fringes and coherent patterns that are beyond the reach of the (generalized) hydrodynamic description. We derive a closed analytic expression for the density profile during the expansion, explicitly highlighting the role played by these interference processes. 
%These findings demonstrate that capturing the dynamics of systems with conformal defects requires a suitably modified semiclassical scattering picture that explicitly accounts for coherent multiple-defect reflections, marking one of the few known instances where non-hydrodynamic effects persist in the thermodynamic limit.
\end{abstract}

\section{Introduction}

Over the past decade, remarkable progress has been achieved in understanding the dynamics of integrable one-dimensional quantum gases, largely driven by the advent of Generalized Hydrodynamics (GHD)~\cite{castro2016emergent,bertini2016transport}. 
One way to interpret the GHD theory, highlighted by many authors, is to regard it as an  emergent classical equation for quasi-particles that propagate ballistically through the system~\cite{castro2016emergent,bertini2016transport,bulchandani2018bethe,doyon2018soliton,caux2019hydrodynamics,bulchandani2019kinetic,gopalakrishnan2018hydrodynamics,hubner2025diffusive}. This picture applies to a wide variety of physical situations and provides a powerful framework to understand also more complex phenomena such as  diffusion~\cite{gopalakrishnan2018hydrodynamics,hubner2025diffusive}, dissipation~\cite{cao2019entanglement,coppola2025exact}, and many more.

One interesting physical situation that has remained relatively unexplored within the framework of GHD is that of localised impurities or defects. The study of impurities has a very rich history in one-dimensional quantum physics, dating back to the Anderson orthogonality catastrophe~\cite{anderson1967infrared} and the Kane-Fisher model~\cite{kane1992transport}. In the latter, a renormalization-group analysis reveals that transport through an impurity in a Luttinger liquid is qualitatively different depending on whether the interactions are attractive or repulsive.
The separation between these two distinct regimes occurs at the free-fermion point, a case that has seen significant progress over the past decades, particularly in the study of entanglement properties at equilibrium, see e.g. \cite{Peschel_2005, PhysRevLett.128.090603, PhysRevLett.102.100502, Kazuhiro_Sakai_2008,https://doi.org/10.1002/andp.201000055,Calabrese_2012,Peschel_2012,Calabrese_2012_2,brehm2015entanglemententropyconformalinterfaces,PhysRevD.95.106008,Capizzi_2022,Capizzi_2022b,Capizzi_2023bb,Mintchev_2021,10.21468/SciPostPhys.11.3.063,Rogerson_2022,Capizzi_2023dd}. 
The out-of-equilibrium description of such systems, however, remains comparatively less developed, largely due to the broad variety of protocols and driving mechanisms that can take a system out of equilibrium. A notable recent development in this direction is Ref.~\cite{rylands2023transport}, which proposes an extended formulation of GHD that incorporates interacting quantum impurity models via the introduction of an impurity collision integral (see Ref. \cite{Collura_2013} for earlier numerical results). Remarkably, this term can be computed exactly by exploiting the integrability of the underlying system.
Significant complementary advances have been made in the context of free-fermionic systems. In particular, Refs.~\cite{Eisler_2012,10.21468/SciPostPhys.14.4.070,Capizzi_2023} analyzed the melting of a domain wall in a free-fermionic chain containing a localized impurity and showed that, for a `conformal' defect ---namely a defect with a transmission coefficient that is independent of the wavelength of the signal, see below---, the full time evolution of the entanglement entropy can be computed exactly.
The importance of this particular initial condition is further underscored by its close connection to a number of earlier studies~\cite{PhysRevB.97.081111,Ljubotina_2019,Scopa2023domainwall,Collura2020domwallmelting,scopa2021exact}, in which domain-wall states were analyzed in the context of transport phenomena within the GHD framework.

Given the central role played by the study of the influence of defects on the dynamics of one-dimensional quantum systems, it is natural to ask how the time evolution is modified when multiple defects are present. This question has so far received no attention at all in the literature. In this work, we address it directly by focusing on the simplest nontrivial case of a domain wall melting through two conformal defects.
Remarkably, we find that even at the level of the density profile, the dynamics cannot be captured by a simple hydrodynamic description. Instead, quantum-interference effects arising from repeated scattering between the defects play a crucial role and must be properly taken into account. Our results therefore demonstrate that the presence of multiple conformal defects leads to qualitatively new dynamical behavior that lies beyond the reach of the (generalized) hydrodynamic approach.

The manuscript is organized as follows.
In Sec. \ref{sec:singledefect} we review known results for one conformal defect.
In Sec. \ref{sec:2defects} we introduce the two defects problem and provide the solution for the propagator. 
In Sec. \ref{sec:density} we derive exact results for the density profile and benchmark them against numerics. 
Finally in Sec. \ref{sec:concl} we draw our conclusions.

\section{Hard-core bosons on a line with a single defect}
\label{sec:singledefect}

\subsection{Hard-core bosons and hydrodynamics of free fermions: no defect}
\label{subsec:nu}
We consider a gas of hard-core bosons hopping on a one-dimensional lattice, described by the Hamiltonian
\be
\hat{H}  = - \frac{1}{2} \sum_{x \in \mathbb{Z}}  ( \hat{\sigma}^+_{x} \hat{\sigma}^-_{x+1} + \hat{\sigma}^+_{x+1} \hat{\sigma}^-_{x}  ) 
\label{eq:XX}
\ee
where $\hat{\sigma}^+_x$ ($\hat{\sigma}^-_x$) is the Pauli operator that creates (destroys) a hard-core boson on site $x$. 
The model has nearest-neighbour hopping, and here we do not consider defects that will be introduced in the next subsection. In spin language, the Hamiltonian  \eqref{eq:XX} is that of the XX spin chain.
It is well-known that, under the Jordan-Wigner mapping
\be    
    \hat{\sigma}_x^+ =  e^{ i \pi \sum_{j = -\infty}^{x-1} \hat{c}_j^\dagger \hat{c}_j} \, \hat{c}_x^\dagger , \qquad
    \hat{\sigma}_x^- = e^{-i \pi \sum_{j = -\infty}^{x-1} \hat{c}_j^\dagger \hat{c}_j}  \,  \hat{c}_x ,
\label{eq:JW_trans}
\ee
where $\hat{c}_x^\dagger$, $\hat{c}_x$, are fermion creation/annihilation operators that obey the canonical anti-commutation relations $\{ \hat{c}_x , \hat{c}_{x'}^\dagger \} = \delta_{x,x'}$, 
the Hamiltonian (\ref{eq:XX}) maps to the one of non-interacting fermions 
\be
\hat{H} = -\frac{1}{2} \sum_{x \in \mathbb{Z}} \left( \hat{c}^\dagger_x \hat{c}_{x+1} +  \hat{c}^\dagger_{x+1} \hat{c}_{x}  \right). 
\label{eq:hoppingH}
\ee
We now give a very brief overview of the hydrodynamic description of the gas of hard-core bosons, which boils down to a particularly simple (non-interacting) version of GHD \cite{bertini2016transport,castro2016emergent}.
Exploiting the Fourier transformation
\be
\begin{split}
\hat{c}(k) = \sum_{x \in \mathbb{Z}} e^{-ikx} \hat{c}_x , \qquad \quad
\hat{c}^\dagger(k) = \sum_{x \in \mathbb{Z}} e^{ikx} \hat{c}_x^\dagger , 
\label{eq:fourier_modes}
\end{split}
\ee
where $\hat{c}(k)$ ($\hat{c}^\dagger(k)$) annihilates (creates) a fermion of momentum $k$, with $k \in \mathbb{R}/2\pi \mathbb{Z}$ (i.e. $k$ is defined modulo $2\pi$), the Hamiltonian (\ref{eq:hoppingH}) is diagonalized as
\be
\hat{H} =  \sum_k \varepsilon(k) \hat{c}^\dagger(k) \hat{c}(k) ,
\ee
with the dispersion relation
\begin{equation}
    \label{eq:dispersion}
\varepsilon(k) = - \cos (k).    
\end{equation}
We now turn to a hydrodynamic description valid at large distances and large times in which the variable $x$ becomes continuous. The state of the gas at time $t$ is described by the phase-space occupation of fermions
\begin{equation}
    \label{eq:dist_nu}
    \nu(x,k,t) \in [0,1],
\end{equation}
where $(x,k) \in \mathbb{R} \times (\mathbb{R}/2\pi \mathbb{Z})$ is a point in phase space. 
%Here the phase-space occupation $\nu$ is normalized such that it is between $0$ and $1$. 
The real-space density of particles at position $x$ is obtained by integrating the phase-space occupation over $k$,
\begin{equation}
    \label{eq:density}
    \rho(x,t) = \int_{-\pi}^\pi  \frac{dk}{2\pi} \nu(x,k,t)  .
\end{equation}
The phase-space occupation evolves in time according to
\begin{equation}
    \label{eq:ghd}
    \partial_t \nu(x,k,t) + v (k) \partial_x \nu(x,k,t)  = 0.
\end{equation}
Here $v(k) = \varepsilon'(k) = \sin(k)$ is the group velocity corresponding to the dispersion relation (\ref{eq:dispersion}).
The function $\nu(x,k)$ can be thought of as the Wigner function of the fermions, that becomes a classical probability distribution (i.e. with no negative values) in the limit of smooth density variations, see for instance the discussions in Refs.~\cite{doyon2017large,ruggiero2019conformal}. The precise definition of the Wigner function for lattice systems can be subtle~\cite{hinarejos2012wigner,hinarejos2015wigner,fagotti2017higher,fagotti2020locally,Essler_2023}; here we simply assume, like in many previous works e.g.~\cite{wendenbaum2013hydro,antal1999transport,Scopa2023domainwall}, that the gas of hard-core bosons at large scales is described by the large-scale occupation (\ref{eq:dist_nu}), which obeys the evolution equation (\ref{eq:ghd}).

Starting from an initial configuration described by  $\nu_0(x,k)$, the solution of the evolution equation~\eqref{eq:ghd} for the occupation function $\nu(x,k,t)$ is given by
\begin{equation}
    \nu(x,k,t) = \nu_0(x - \sin(k)t,k).
\end{equation}
%with group velocity resulting from the lattice dispersion relation of the free fermions:
%\begin{equation}
%    v(k) = \frac{{\rm d} \varepsilon(k)}{{\rm d}k} =  - \frac{{\rm d}}{{\rm d}k} \cos(k) = \sin(k).
%    \label{eq:disp_rel}
%\end{equation}
We stress that the description of the dynamics of the gas in terms of the occupation ratio (\ref{eq:dist_nu}) and the evolution equation (\ref{eq:ghd}) is entirely classical. It does capture some quantities, for instance the density profile (\ref{eq:density}), in the limit of large time and large distances. However, it does not give access to other quantities which contain correlations resulting from the `quantum' nature of the system. Obtaining quantities like the one-particle density matrix and entanglement entropy is beyond the scope of this work and would require the requantisation of the theory, as done in previous works \cite{takács2024quasicondensation,dubail2017conformal,ruggiero2019conformal,ruggiero2020quantum,Collura2020domwallmelting,ruggiero2021quantum,scopa2021exact,scopa2022exact,Scopa_2023}.
Instead, in this paper we focus on an intriguing class of post-quench free-fermion dynamics in which the presence of conformal defects~\cite{Peschel_2012,10.21468/SciPostPhys.14.4.070} gives rise to genuinely quantum behavior already at the level of the density profile, leading to a breakdown of the standard hydrodynamic description.

\subsection{Hard-core bosons with a defect}

\begin{figure}
    \centering
    \includegraphics[width =0.8\textwidth]{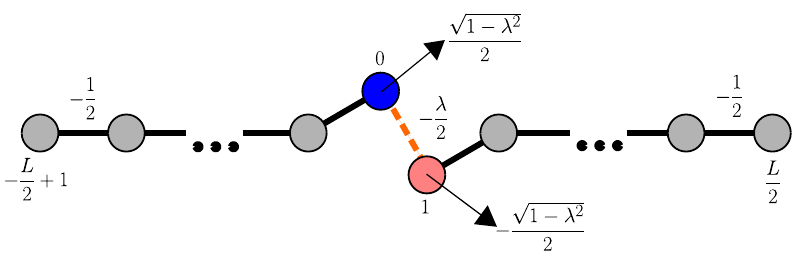}
    \caption{Sketch of the nearest-neighbor hopping  Hamiltonian $\hat{H}_\lambda$ of Ref.~\cite{Capizzi_2023},  given by Eq.~\eqref{eq:singledefect}, parameterised by the strength of the impurity $\lambda \in [0,1]$.}
    \label{fig:defect}
\end{figure}
The dynamics of lattice hard-core bosons or free fermions with a single defect was investigated  in previous works \cite{Peschel_2012,10.21468/SciPostPhys.14.4.070,Capizzi_2023}. In particular, an emergent quasi-particle picture was derived, which allows the calculation of the density profile and other quantities such as the bipartite entanglement entropy to leading order in time. The model considered in Ref.~\cite{Capizzi_2023} is the hard-core boson Hamiltonian with a `conformal' defect,
    \begin{equation}
        \hat{H}_\lambda = \sum_{x \in \mathbb{Z}} h_{x,y} \hat{\sigma}_x^+ \hat{\sigma}_{y}^-
        \label{eq:hoppingH}
    \end{equation}
with $h_{x,y}=0$ if $|x-y| \geq 2$, $h_{x,x+1} = h_{x+1,x} = -\frac{1}{2}$ for $x \neq 0,1$, and
\begin{equation}
    h_{0,1} = h_{1,0} = -\frac{\lambda}{2} \quad h_{0,0} = - h_{1,1} = \frac{1}{2}\sqrt{1 - \lambda^2},
    \label{eq:singledefect}
\end{equation}
with a real parameter $\lambda \in \left[0,1\right]$ representing the strength of the defect.  
An illustration of this model is shown in Fig.~\ref{fig:defect}. This Hamiltonian has a weak link between sites $x= 0$ and $x= 1$, as well as a staggered on-site potential. Here, we also introduce a finite system size $L$ as well as periodic boundary conditions. Later in the calculation we will take the thermodynamic limit $L \rightarrow \infty$. Even though we introduced an impurity, or a weak link, into the system, the Jordan-Wigner transformation mapping the hard-core bosons to free fermions (discussed above in Eq. \eqref{eq:JW_trans}) is still applicable, since the Hamiltonian \eqref{eq:hoppingH} still consists only of nearest-neighbour or on-site quadratic terms. Thus the Hamiltonian of the system with size $L$ (chosen to be even for convenience) and periodic boundary conditions is equivalent to
\begin{equation}
\hat{H}_{\lambda} = \sum_{-L/2 +1  \leq x,y \leq L/2} h_{x,y} \hat{c}^\dagger_x \hat{c}_y,
\label{eq:hoppingH2}
\end{equation}
with the same hopping terms as in Eqs.~\eqref{eq:hoppingH}-\eqref{eq:singledefect}, written in terms of fermionic annihilation/creation operators $\hat{c}_i$/$\hat{c}_j^\dagger$. The boundary conditions of the fermions are periodic/anti-periodic depending on whether the parity of the total number of fermions is odd/even respectively. The defect is located at positions $x = 0,1$ given by Eq. \eqref{eq:singledefect}. We also introduced the notation $\hat{H}_\lambda$ to denote a Hamiltonian with one conformal defect characterized by the parameter $\lambda$. To understand the effect the conformal defect has on the many-particle dynamics we start by diagonalizing $\hat{H}_\lambda$ in the single-particle sector. We write $\Phi_k(x)$ for the eigenstates of the $L \times L$ matrix $h$ in Eq.~(\ref{eq:hoppingH2}),
\begin{equation}
    h_{x,y} \Phi_k(y) = \varepsilon_k \Phi_k(x),
    \label{eq:one_defect_eignespec}
\end{equation}
where $\varepsilon_k$ is the single-particle eigenvalue and $k$ is a quantum number that labels the eigenstates. The many-body eigenstates of the Hamiltonian $\hat{H}_{\lambda}$ can then be obtained as Slater determinants of $\Phi_k(x)$ for a set of the occupied quantum numbers $k$. Coming back to the single-particle eigenstate $\Phi_k(x)$, we use the ansatz
\begin{equation}
    \Phi_k(x) = \begin{cases}
        Ae^{ikx} + Be^{-ikx}, \qquad \text{if}  \qquad x \in [ -\frac{L}{2} + 1, 0] \\
        Ce^{ikx} + De^{-ikx}, \qquad \text{if} \qquad x \in [ 1, \frac{L}{2}],
    \end{cases}
    \label{eq:1defect_ansatz}
\end{equation}
where we intentionally omit any $k$ dependence of the amplitudes $A,B,C$ and $D$ as we show later, that due to the nature of the defects, these amplitudes are independent of the quasimomenta.
Plugging this ansatz into Eq.~(\ref{eq:one_defect_eignespec}), and looking first at points $x \notin \{ 0,1, -\frac{L}{2}+1, \frac{L}{2} \}$, if the ansatz (\ref{eq:1defect_ansatz}) is an eigenstate, then the associated energy is
\begin{equation}
    \label{eq:dispers}
    \varepsilon_k = \varepsilon(k) = - \cos(k) .
\end{equation}
Then, looking at the constraint imposed by Eq.~(\ref{eq:one_defect_eignespec}) on $\Phi_k(x)$ at $x=0$ and $x=1$, the amplitudes $A,B$ are related to the amplitudes $C,D$ through
\begin{equation}
\label{eq:1defectrelation}
    \begin{pmatrix}
        A \\ 
        B
    \end{pmatrix} = \mathbf{M}_\lambda \begin{pmatrix}
        C \\
        D
    \end{pmatrix},
\end{equation}
where 
\begin{equation}
    \mathbf{M}_\lambda = \frac{1}{\lambda} \begin{pmatrix}
        1 & -\sqrt{1 - \lambda^2} e^{-ik} \\
        -\sqrt{1 - \lambda^2} e^{ik} & 1 
    \end{pmatrix} .
\end{equation}
Similarly, the periodic boundary conditions at $x=-\frac{L}{2}+1, \frac{L}{2}$,  
\begin{equation}
    \Phi_k(x) = \Phi_k(x + L),
\end{equation}
give another relation between amplitudes $A,B$ and $C,D$,
\begin{equation}
\label{eq:1defect_cond2}
    \begin{pmatrix}
        C \\ 
        D
    \end{pmatrix} = \mathbf{T}_L \begin{pmatrix}
        A \\
        B
    \end{pmatrix},
\end{equation}
where the matrix $\mathbf{T}_L$ is
\begin{equation}
    \mathbf{T}_L =  \begin{pmatrix}
        e^{-ikL} & 0  \\
        0 & e^{ikL} 
    \end{pmatrix}.
\end{equation}
Combining the two conditions~\eqref{eq:1defectrelation} and \eqref{eq:1defect_cond2}, we get a system of equations for the amplitudes $A$ and $B$ in the form
\begin{equation}
    \begin{pmatrix}
        A \\ 
        B
    \end{pmatrix} = \mathbf{M}_\lambda \mathbf{T}_L \begin{pmatrix}
        A \\
        B
    \end{pmatrix}.
\end{equation}
This system of equations has a non-trivial solution if and only if
\begin{equation}
    {\rm Det} \left[ \mathbf{1} - \mathbf{M}_\lambda \mathbf{T}_L \right] = 0.
\end{equation}
This is a quantization condition, that constrains the quantum number $k$ to a specific set of allowed values that depends on the system size $L$ and on the defect parameter $\lambda$. Organizing the terms that explicitly depend on $k$ on the l.h.s of the equation, we find a condition on $k$ in terms of the parameter $\lambda$ of the impurity
\begin{equation}
    \lambda = \cos(kL),
\end{equation}
resulting in the quantization condition
\begin{equation}
    k = \frac{\arccos{\lambda}}{L} + \frac{2 \pi n}{L}, \qquad \text{for } n = 1,...,L.
    \label{eq:kcond_1}
\end{equation}
This condition can also be rewritten as
\begin{equation}
    \label{eq:qcond_1_alternate}
    e^{ikL} = \lambda + i\sqrt{1 - \lambda^2}.
\end{equation}
Since we found $n$ independent solutions and the matrix $h$ is of size $n\times n$, our ansatz exhausts all possible single-particle eigenstates. After properly normalizing them, all the eigenstates are given by
\begin{equation}
    \Phi_k(x) = \frac{1}{\sqrt{2L}}\begin{cases}
       e^{ikx} - ie^{ik(L+1)}e^{-ikx} , \quad {\rm for } \quad x \in [1,\frac{L}{2}] \\
     e^{ikL}e^{ikx} - ie^{-ik(x-1)} , \quad {\rm for } \quad x \in [-\frac{L}{2}+1,0]
    \end{cases} 
    \label{eq:1defectwavefunc}
\end{equation}
with energy (\ref{eq:dispers}) and where the wavevector $k$ is in the set (\ref{eq:kcond_1}).

\subsection{The propagator}

Our next goal is to revisit the quasiparticle picture introduced in Ref.~\cite{Capizzi_2023} from the perspective of the fermionic propagator in the presence of a conformal defect. This analysis leads us to the explicit expression~(\ref{eq:prop_final1}) for the propagator that was first obtained in ~\cite{10.21468/SciPostPhys.14.4.070}. We show that the quasiparticle picture of Ref.~\cite{Capizzi_2023} emerges naturally from this result, and we recover their findings for the particle density profile. The derivation of the propagator in the presence of a single defect provides the foundation for addressing the more involved case of two defects, which we study in Section~\ref{sec:2defects}.

The fermionic propagator $K_t(x,y)$ is the kernel that relates the Heisenberg-picture annihilation operator $\hat{c}_x(t)$ to the operators $\hat{c}_y(0) = \hat{c}_y$ at time $t=0$,
\be
\hat{c}_x(t)  = \sum_{y = -L/2+1}^{L/2} K_t(x,y) \hat{c}_y(0).
\label{eq:propagator}
\ee
We recall that the Heisenberg-picture fermion annihilation operator evolves according to
\begin{equation}
    \frac{{\rm d}}{{\rm d} t}  \hat{c}_x(t) = i \left[ \hat{H}_\lambda, \hat{c}_x(t)\right].
    \label{eq:op_time_ev}
\end{equation}

To compute the propagator $K_t(x,y)$, we use the completeness relation in the single-particle eigenbasis $\Phi_k(x)$ of the Hamiltonian $\hat{H}_\lambda$, which allows us to write
\begin{equation}
    K_t(x,y) = \sum_{k} e^{it\varepsilon(k)} \Phi_k(x) \Phi_k^*(y),
    \label{eq:propagator_def}
\end{equation}
where the sum over $k$ goes over the set of quasimomenta fixed by the quantisation condition \eqref{eq:kcond_1}. For concreteness, let us focus first on the case $x,y \geq 1$. Plugging the eigenstates (\ref{eq:1defectwavefunc}) into Eq. \eqref{eq:propagator_def}, we arrive at
\begin{eqnarray*}
    (x,y \geq 1) \qquad  K_t(x,y) &=& \frac{1}{2L} \sum_{k}   e^{-i[k(y-x) - t\varepsilon(k)]} + e^{-i[k(x-y) - t\varepsilon(k)]}  \nonumber \\ 
    && + \frac{1}{2L} \sum_{k} ie^{-i[k(L+1-x-y)  - t\varepsilon(k)]} - ie^{-i[k(x+y-L-1) - t\varepsilon(k)]}.
\end{eqnarray*}
To simplify the system size dependence in the exponents, we employ the quantization condition in Eq. \eqref{eq:qcond_1_alternate} that relates $e^{ikL}$ to the parameter $\lambda$. Finally, we take the thermodynamic limit $L \rightarrow \infty$ and replace the discrete sum over $k$ by an integral, $ \sum_k \rightarrow \int_{0}^{2\pi} \frac{L}{2 \pi} {\rm d }k $.
 After some simple manipulations, we obtain the propagator for a single defect in the thermodynamic limit,
\be
    \label{eq:Kt_30_1}
(x,y \geq 1) \qquad  K_t(x,y) = K_1 (x,y,t) 
 =   (-i)^{x-y} J_{x-y}(t) + \sqrt{1 - \lambda^2}(-i)^{x+y-1} J_{x+y-1}(t)
\ee
with $  J_n(t) = \frac{1}{2 \pi} \int_{-\pi}^\pi e^{i (n (-\pi/2 + k) + t \cos(k))} {\rm d}k $ the $n^{\rm th}$ Bessel function of the first kind. The other cases $(x \leq 0, y \geq 1)$ etc. can be treated in the same way. The final result is better summarized in a table,
\begin{equation}
\begin{tabular}{|c|c|c|}
    \hline 
    $K_t(x,y)$ &  $y \leq 0$ &  $y \geq 1$ \\
    \hline
    $x \leq 0$ & $K_1(x,y,t)$ & $K_2(x,y,t)$\\
    \hline
    $x \geq 1$ & $K_2(x,y,t)$ & $K_1(x,y,t)$  \\
    \hline 
\end{tabular}
\label{eq:prop_final1}
\end{equation}
where $K_1(x,y,t)$ is defined above, and 
\begin{equation}
    K_2(x,y,t) = \lambda (-i)^{x-y}  J_{x-y}(t) .
    \label{eq:K1_K2}
\end{equation}

\subsection{Quasi-particle interpretation of the propagator (\ref{eq:prop_final1})}
\label{section:Quasi_part_prop}

\begin{figure}
    \centering
    \includegraphics[width=0.49\linewidth]{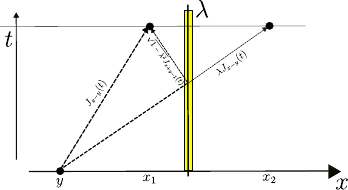}
    \includegraphics[width = 0.49\linewidth]{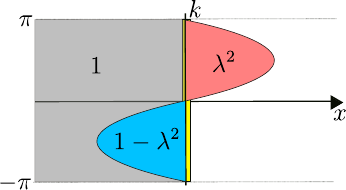}
    \caption{Left: Schematic drawing of a particle propagating from $y$ to $x_\tau$ ($- x_\tau$) at time $\tau$ interpreted in the quasi particle picture. A particle initially at position $y < 0$ with quasimomentum $k$ and velocity $v(k) = \sin(k)$ arriving at the defect at $x = 0$ gets reflected (transmitted) with an amplitude given by $R(\lambda)$ ($T(\lambda)$). Right: the emerging hydrodynamic picture. Considering a domain wall  \eqref{eq:DW_initstate} and evolving it according to Eq. \eqref{eq:timev_unitary} results in the  phase-space picture given by hydrodynamic description. }
    \label{fig:onedefect_quasiparticle}
\end{figure}

The above propagator has a physical interpretation which will be very important in what follows. To see this, let us imagine that the particle is initially in a semi-classical wavepacket around position $(x_0,k_0)$ in phase-space, corresponding to a wavefunction of the form
\begin{equation}
    \psi_{x_0,k_0} (y) \, \sim \, e^{i k_0 y} e^{- ( y- x_0 )^2/(2 \ell^2) }  .
\end{equation}
Here the size $\ell$ of the wavepacket is chosen such that $1 \ll \ell \ll |x_0|$, 
so that it extends on many lattice sites around $x_0$ and is far from the defect at $x=0$. We assume that $x_0<0$ and $-x_0  \gg \ell $ so that the semiclassical wavepacket is on the left of the defect. Then at times $t>0$ that are not too long so that $t \ll \ell^2$, the dispersion of the wavepacket remains small, and we have 
\begin{multline}
    \label{eq:wp_approx}
    \sum_{y\in \mathbb{Z}} K^*_t(x,y) \psi_{x_0,k_0} (y) \, \simeq \, \\ e^{-i \varepsilon(k_0) t } \times \left\{ \begin{array}{ccc}
         \psi_{x_0 + v(k_0) t ,k_0} (x) &\quad {\rm if}& v(k_0) t < |x_0| \\
       \lambda \psi_{x_0 + v(k_0) t ,k_0} (x) + \sqrt{1-\lambda^2}  \psi_{-x_0 - v(k_0) t , -k_0} (x)  &\quad {\rm if}& v(k_0) t > |x_0|   ,
    \end{array} \right. 
\end{multline}
which means that the wavepacket evolves ballistically at velocity $v(k_0) = \varepsilon'(k_0)$ until it hits the defect, and then it splits into two wavepackets, one transmitted (with wavevector $+k_0$) and one reflected (wavevector $-k_0$), with corresponding transmission/reflection amplitudes $\sqrt{1-\lambda^2}$ and $\lambda$ respectively.

\vspace{0.5cm}

Eq.~(\ref{eq:wp_approx}) can be derived from Eq.~(\ref{eq:prop_final1}) as follows. For simplicity we focus on the case $v(k_0) t < |x_0|$; the case $v(k_0) t > |x_0|$ follows from a similar calculation. In the former case, what matters is the first term in the r.h.s of (\ref{eq:Kt_30_1}),
\begin{eqnarray}
    \label{eq:propwp}
\sum_{y\in \mathbb{Z}} i^{x-y} J_{x-y}(t) \psi_{x_0,k_0} (y) & = &  \sum_{y \in \mathbb{Z}} \int_{-\pi}^\pi \frac{dq}{2\pi} e^{i q (x-y) - it \varepsilon(q) }   e^{i k_0 y} e^{- ( y- x_0 )^2/(2 \ell^2) } .
\end{eqnarray}
Using the Poisson resummation formula $\sum_{y\in \mathbb{Z}} \rightarrow \sum_{m \in \mathbb{Z}} \int dy \,  e^{i 2 \pi m y}$, the sum becomes 
\begin{eqnarray*}
    \sum_{y \in \mathbb{Z}} e^{i q (x-y) - it \varepsilon(q) }  e^{i k_0 y} e^{- ( y- x_0 )^2/(2 \ell^2) }  & = & 
   e^{i q x - it \varepsilon(q)} \sum_{m \in \mathbb{Z}} \int dy \, e^{i  2\pi m y}   e^{i (k_0-q) y} e^{- ( y- x_0 )^2/(2 \ell^2) }  \\ &=&  e^{i q  x - it \varepsilon(q) } \sum_{m \in \mathbb{Z}}   \sqrt{2\pi} \ell\, e^{ i  (k_0 + 2\pi m - q) x_0} e^{ - \frac{(k_0 + 2\pi m-q)^2 \ell^2}{2} }  \\
    &\underset{\ell \gg 1}{\simeq} & \sqrt{2\pi} \ell \, e^{i q  x - it \varepsilon(q) }   e^{ i (k_0-q) x_0 } e^{ - \frac{(k_0-q)^2 \ell^2}{2} }  \, .
\end{eqnarray*}
From the first to the second line we evaluated the Gaussian integral, and from the second to the third line we used the fact that the sum is dominated by the $m=0$ term when $\ell \gg 1$.
Plugging this into Eq.~(\ref{eq:propwp}), one is left with an integral that can be evaluated in the saddle-point approximation,
\begin{eqnarray}
    \label{eq:saddle-point_gauss}
\nonumber \sum_{y\in \mathbb{Z}} i^{x-y} J_{x-y}(t) \psi_{x_0,k_0} (y) & = &  \sqrt{2\pi} \ell  \int_{-\pi}^\pi \frac{dq}{2\pi}  e^{i q  x - it \varepsilon(q) }  e^{ i (k_0-q) x_0} e^{ - \frac{(k_0-q)^2 \ell^2}{2} }  \\
\nonumber    &  \simeq &  e^{ i k_0 x_0} e^{i q_* (x-x_0) - i t \varepsilon(q_*) }  e^{ - \frac{(k_0-q_*)^2  \ell^2}{2} }  \\ 
\nonumber    && \times \sqrt{2\pi} \ell \int_{-\infty}^\infty \frac{dq}{2\pi}  e^{ - \frac{i t  \varepsilon''(q_*) + \ell^2}{2} (q-q_*)^2   }  \\
& = & e^{ i k_0 x_0} e^{i q_* (x-x_0) - i t \varepsilon(q_*) }  e^{ - \frac{(k_0-q_*)^2  \ell^2}{2} } \left( 1 + i \frac{t \varepsilon''(q_*)}{\ell^2} \right)^{-1/2} .
\end{eqnarray}
Here $q_*$ is the solution of the saddle-point equation
\begin{equation}
    \label{eq:saddle-point}
    x-x_0 - t \, v(q_*) + i (q_*-k_0) \ell^2 = 0 \, .
\end{equation}
For $|x-x_0|, t \ll \ell^2$ we see that $q_* \simeq k_0 + i  \frac{ x-x_0 - t v(k_0) }{\ell^2} $ to first order in $|x-x_0|/\ell^2$ and $t/\ell^2$. %Expanding (\ref{eq:saddle-point}) to first order we get $x-x_0 - t v(k_0) - t \varepsilon''(k_0) (q_* - k_0) + i (q_* - k_0)\ell^2 = 0$, so we see that the saddle point is
%\begin{equation}
%    q_*  =  k_0 +  i  \frac{ x-x_0 - t v(k_0) }{ \ell^2 + i t \varepsilon''(k_0)   }  + \dots
%\end{equation}
Plugging this into (\ref{eq:saddle-point_gauss}), we arrive at
\begin{eqnarray}
  \nonumber   \sum_{y\in \mathbb{Z}} i^{x-y} J_{x-y}(t) \psi_{x_0,k_0} (y)  & \simeq &  e^{ i k_0 x_0} e^{i k_0 (x-x_0) - i t \varepsilon(k_0) - i t v(k_0) (q_* - k_0) + i (q_*-k_0) (x-x_0) }  e^{ - \frac{(k_0-q_*)^2  \ell^2}{2} } \\
\nonumber  & =  &  e^{- i t \varepsilon(k_0)} e^{ i k_0 x}  e^{ -\frac{(x-x_0 - v(k_0) t)^2 }{2 \ell^2} }  \\
  &= &  e^{- i t \varepsilon(k_0) }  \, \psi_{x_0 + v(k_0) t,k_0}(x) ,
\end{eqnarray}
which is Eq.~(\ref{eq:wp_approx}) as claimed. Note that when using (\ref{eq:saddle-point_gauss}) we used $\left( 1 + i \frac{t \varepsilon''(q_*)}{\ell^2} \right)^{-1/2} \simeq 1$ for $t/\ell^2 \ll 1$.

\vspace{0.5cm}

Thus, we see that the propagator (\ref{eq:prop_final1}) has the aforementioned semi-classical interpretation, illustrated in Fig.~\ref{fig:onedefect_quasiparticle} (Left): fermionic particles travel ballistically with velocity $v(k)$ until they reach the defect at position $x = 0$. At the defect they scatter and are transmitted or reflected with probability $T(\lambda) = \lambda^2$ or $R(\lambda) = 1 - \lambda^2$ respectivelly, and after that they move again ballistically.

Coming back to the description of the gas in terms of the large-scale phase-space occupation $\nu(x,k,t)$ discussed in Section~\ref{subsec:nu}, we see that the evolution equation
\begin{equation}
    \label{eq:ghd_bc}
\partial_t \nu(x,k,t)  + \sin k \, \partial_x \nu (x,k,t) =  0
\end{equation}
that encodes the ballistic propagation remains valid away from the defect, i.e. for $x\neq 0$. In that description, the defect enters as a boundary condition at $x=0$, relating the quasi-particle occupation on the left ($x=0^-$) and right ($x=0^+$):
\begin{equation}    
    \label{eq:bc_nu_defect}
\left\{ \begin{array}{rclccl}
    \nu (0^+ , k , t ) &=&  \lambda^2 \nu (0^- , k , t ) + (1-\lambda^2) \nu (0^+ , -k , t )  & \qquad & {\rm if} & v(k) >0 \\ 
    \nu (0^- , k , t ) &=&  \lambda^2 \nu (0^+ , k , t ) + (1-\lambda^2) \nu (0^- , -k , t )  & \qquad & {\rm if} & v(k) < 0 \, .
\end{array} \right.
\end{equation}

\subsection{Application: the density profile after 
domain-wall quench through the defect}

In Ref. \cite{Capizzi_2023} the density profile of a system of free spinless fermions scattering on a conformal defect of the form \eqref{eq:singledefect} was calculated after a quench from the initial condition 
\begin{equation}
    \ket{{\rm DW}} = \bigotimes_{x = -L/2+1 }^0 \ket{1} \bigotimes_{x = 1 }^{L/2} \ket{0},
    \label{eq:DW_initstate}
\end{equation}
where $\ket{0}$ and $\ket{1}$ are the eigenstates of the fermionic number operator $\hat{\rho}(x) = \hat{c}_x^\dagger \hat{c}_x$ with eigenvalues $0,1$. The system evolves according to the unitary time evolution
\begin{equation}
    \ket{\Psi(t)} = e^{- i \hat{H}_\lambda t} \ket{{\rm DW}}.
    \label{eq:timev_unitary}
\end{equation}
The initial occupation function is given by
\be
\nu_0(x,k) = \begin{cases}
    1 \quad \text{ if } \quad x \leq 0 , \\
    0 \quad \text{ otherwise}.
\end{cases}
\ee
Accounting for the scattering on the defect, the solution to the hydrodynamic equation (\ref{eq:ghd_bc})-(\ref{eq:bc_nu_defect}) is 
\be
\nu(x,k,t) = \begin{cases}
    1 & \quad \text{ if } \quad x \leq \min[\sin(k)t,0] \\
    1 - \lambda^2 & \quad \text{ if } \quad  \min[\sin(k)t,0] < x \leq 0 \\
    \lambda^2 & \quad \text{ if } \quad  0 < x \leq \max[\sin(k)t,0] \\
    0 & \quad \text{ otherwise},
\end{cases}
\label{eq:domain_wall_occupation_func}
\ee
as illustrated in Fig. \ref{fig:onedefect_quasiparticle}. Using Eq.~(\ref{eq:density}) this gives the hydrodynamic density profile 
\begin{eqnarray}
    \label{eq:res_dens1}
    \rho(x,t) = \begin{cases}
        1 - \frac{\lambda^2}{\pi} \arccos\left(\frac{-x}{t}\right), &\quad { \rm for } -t \leq x \\
        1, &\quad {\rm otherwise}
    \end{cases}
\end{eqnarray}
for $x \leq 0$ and 
\begin{eqnarray}
    \label{eq:res_dens2}
    \rho(x,t) = 
        \frac{\lambda^2}{\pi} \arccos\left(\frac{x}{t}\right)
\end{eqnarray}
for $1 \leq x$, which is the main result of Ref.~\cite{Capizzi_2023}.

Alternatively, this result can be derived directly from the propagator $K_t(x,y)$. Indeed, in the time evolution of the expectation value of the density is
\begin{eqnarray}
    \rho_t(x) = \bra{{\rm DW}} \hat{c}^\dagger_x(t) \hat{c}_x(t) \ket{{\rm DW}},
\end{eqnarray}
where the time dependence of $\hat{c}^\dagger_x(t)$/$\hat{c}_x(t)$ is governed by Eq. \eqref{eq:op_time_ev}. Using the definitions \eqref{eq:propagator_def} and  \eqref{eq:DW_initstate}, the fermion density can be expressed as
\begin{align}
    &\rho(x,t) = \sum_{y\leq 0} K_t(x,y) K^*_t(x,y)  %\notag \\&
    = \begin{cases} \displaystyle
    \sum_{y \leq 0} J^2_{x-y}(t) + (1 - \lambda^2) J^2_{x+y-1}(t), & \quad \text{for} \quad x \leq 0 \\
    \displaystyle
    \sum_{y \leq 0} \lambda^2  J_{x-y}(t) & \quad\text{for} \quad 1 \leq x.
    \end{cases}
    \label{eq:1def_dens1}
\end{align}
Note the absence of cross terms arising from the different contributions of the propagator in Eqs. \eqref{eq:prop_final1}, as these terms cancel exactly. Taking the hydrodynamic (Euler) limit and writing $\sum_{y \leq 0} \rightarrow \int d{\rm y}$, the terms in the integral can be evaluated using the saddle point approximation of $J_n(t)$
\begin{equation}
    J_n(t) \approx \sqrt{\frac{2}{\pi \sqrt{t^2 - n^2}}} \cos\left( \sqrt{t^2 - n^2} - n\arccos{\frac{n}{t}} - \frac{\pi}{4} \right)
    \label{eq:Bessel_straddle}
\end{equation}for $t \rightarrow \infty $ and $\frac{n}{t}$ fixed, giving again the density profile (\ref{eq:res_dens1})-(\ref{eq:res_dens2}).

\section{The case with two defects}
\label{sec:2defects}

%In Section~\ref{sec:singledefect}, we have revisited the setup of the hard-core boson gas with a weak link, focusing on the propagator $K_t(x,y)$ and on its interpretation in terms of quasiparticle reflection and transmission through the defect, and we have recovered the results of Ref.~\cite{Capizzi_2023}. The large-scale hydrodynamic picture corresponds to the one of Fig.~\ref{fig:onedefect_quasiparticle} where each quasiparticle travels ballistically through the system and is either reflected or transmitted by the defect with probability $1-\lambda^2$ and $\lambda^2$ respectively. Note that this emergent large-scale behavior is entirely classical.

\begin{figure}
    \centering
    \includegraphics[width = \linewidth ]{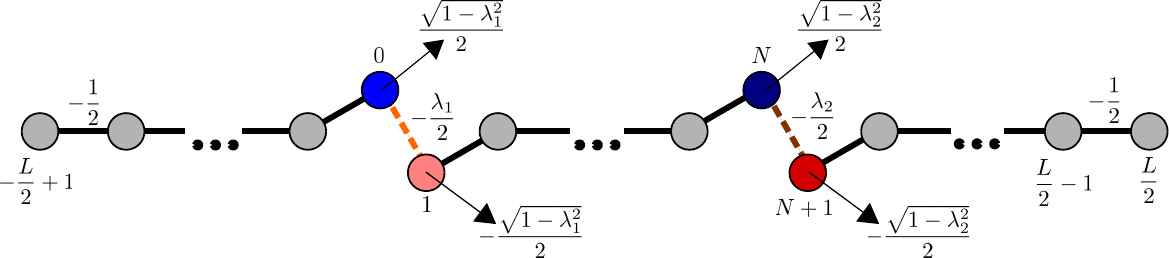}
    \caption{Sketch of the system with Hamiltonian $\hat{H}_{\lambda_1, \lambda_2}$ given by Eqs. \eqref{eq:hoppingH2} and \eqref{eq:2defects}, parametrised by the strength of the two impurities $\lambda_1,\lambda_2 \in \left[0,1\right]$, that introduce a staggered onsite potential on site $0,1$ and $N,N+1$ as well as a different hopping terms. }
    \label{fig:2defects}
\end{figure}

In this section we turn to the main setup of interest in this paper, namely the case of two defects separated by a distance $N$, as illustrated in Fig.~\ref{fig:2defects}. As anticipated, the presence of the second defect modifies the dynamics in a nontrivial way. Interferences between repeatedly reflected particles give rise to genuinely quantum effects that are already visible at the level of the density profile, and cannot be captured by a hydrodynamic equation of the form (\ref{eq:ghd_bc}). As we discuss below, a quasiparticle interpretation can nevertheless be recovered by properly accounting for these interference contributions.
%We conclude this section by looking at the evolution of the density profile after a quench from the domain wall initial state defined in Eq. \eqref{eq:DW_initstate} and compare our findings to free fermion numerics.

\subsection{Model with two defects of strengths \texorpdfstring{$\lambda_1$, $\lambda_2$}{lambda1,lambda2}}
The Hamiltonian of interest $\hat{H}_{\lambda_1, \lambda_2}$ is given by Eq.~\eqref{eq:hoppingH2} with hopping $h_{x,x+1} = h_{x+1,x} = -\frac{1}{2}$ for $x\notin \{ 0,1, N,N+1 \}$, and two weak links of strengths $\lambda_1$ and $\lambda_2$ at $x = 0,1$ and at $x = N, N+1$ respectively,
\begin{eqnarray}
        \label{eq:2defects}
    h_{0,1} = h_{1,0} = -\frac{\lambda_1}{2}\quad \text{and} \quad h_{N,N+1} = h_{N+1,N} = -\frac{\lambda_2}{2}, \quad  \text{with} \nonumber \\
    \quad h_{0,0} = - h_{1,1} = \frac{1}{2}\sqrt{1 - \lambda_1^2} \quad \text{and} \quad h_{N,N} = - h_{N+1,N+1} = \frac{1}{2}\sqrt{1 - \lambda_2^2} \, .
\end{eqnarray}
We take both parameters $\lambda_1,\lambda_2 \in \left[0, 1 \right]$.
To obtain the propagator and the corresponding density profile, we follow the same strategy as in the single-defect case. Specifically, we first determine the single-particle eigenstates of the Hamiltonian $\hat{H}_{\lambda_1,\lambda_2}$ under periodic boundary conditions, identifying the sites $x = -\frac{L}{2}+1$ and $x = \frac{L}{2}$, where $L$ is taken sufficiently large such that $L > N/2$.

To find the eigenstates $\Psi_{k}(x)$ of the $L\times L$ hermitian matrix $h$, we start with the ansatz
\begin{equation}
    \label{eq:psik_2defects_1}
     \Psi_k(x) \propto \begin{cases}
         A e^{ikx} + B e^{-ik(x-1)}& \qquad \text{for } x \in [-L/2 +1,0 ] \\
         C e^{ikx} + D e^{-ik(x-1)}& \qquad \text{for } x \in [0,N+1 ] \\
          e^{ikx} +  F e^{-ik(x-1)} &\qquad \text{for } x \in [N+1,L/2 ] ,
     \end{cases}
\end{equation}
for an eigenstate with energy (\ref{eq:dispers}). 
The relation between the amplitudes $A,B,C,\dots$ is obtained by writing the constraints $\Psi_k(x)$ has to satisfy at the two defects, and the constraints arising from the periodic boundary condition. The resulting system of equations is
\begin{subequations}
\begin{equation}  
    \begin{pmatrix}
        1 \\
        F
    \end{pmatrix} = \begin{bmatrix}
        e^{-ikL} & 0 \\
        0 & e^{ikL}
    \end{bmatrix}\begin{pmatrix}
      A \\
      B
    \end{pmatrix}= {\bf T}_L \begin{pmatrix}
      A \\
      B
    \end{pmatrix},
\end{equation}
coming from the periodic boundary condition of $\Psi_k(x) = \Psi_k(x+L)$,
\begin{equation}
  \begin{pmatrix}
        A \\
        B
    \end{pmatrix} = \frac{1}{\lambda_1}\begin{bmatrix}
        1 & -\sqrt{1 - \lambda_1^2} \\
        -\sqrt{1 - \lambda_1^2} & 1
    \end{bmatrix}\begin{pmatrix}
      C \\
      D
    \end{pmatrix}= {\bf M}_{\lambda_1} \begin{pmatrix}
      C \\
      D
    \end{pmatrix}
\end{equation}
coming from the defect located at $x = 0,1$ and 
\begin{equation}
    \begin{pmatrix}
        C \\
        D
    \end{pmatrix} = \frac{1}{\lambda_2}\begin{bmatrix}
        1 & -\sqrt{1 - \lambda_2^2}e^{-i2Nk} \\
        -\sqrt{1 - \lambda_2^2}e^{i2Nk}  & 1
    \end{bmatrix}\begin{pmatrix}
      1 \\
      F
    \end{pmatrix}=  {\bf T}_N {\bf M}_{\lambda_2} {\bf T}_N^{-1} \begin{pmatrix}
      1 \\
      F
    \end{pmatrix},
\end{equation}
\label{eq:matrices_M}
\end{subequations}
from the defect located at $x = N,N+1$. This system of equations has a solution if and only if
\begin{equation}
    {\rm Det}\left[ \mathbf{1} - {\bf T}_L {\bf M}_{\lambda_1}  {\bf T}_N {\bf M}_{\lambda_2} {\bf T}_N^{-1}  \right] = 0,
\end{equation}
which provides us with a quantisation condition for $k$ in the form
\begin{eqnarray}
        \lambda_1\lambda_2 = \cos(kL) + \sqrt{1 - \lambda_1^2}\sqrt{1 - \lambda_2^2}\cos(kL - 2kN).
    \label{eq:quant_cond_2def_1}
\end{eqnarray}
Alternatively, the quantisation condition can be expressed as
\begin{equation}
    e^{ik_{\pm}L} = \frac{\lambda_1\lambda_2 \pm i \sqrt{2 - \lambda_1^2  - \lambda_2^2 + 2\sqrt{1 - \lambda_1^2}\sqrt{1 - \lambda_2^2}\cos(k_{\pm} 2 N)}}{1 + \sqrt{1 - \lambda_1^2}\sqrt{1 - \lambda_2^2}e^{-ik_{\pm}2N}},
    \label{eq:quant_cond_2def_4}
\end{equation}
where we see the appearance of two sets of solutions, depending on the root that we pick, with each root containing $\frac{L}{2}$ solutions (assuming even $L$). The amplitudes can then be written as 
\begin{equation}
\begin{matrix}
    F_{\pm} = \frac{\frac{\lambda_2}{\lambda_1}
    \sqrt{1-\lambda_1^2}z_\pm + \sqrt{1 - \lambda_2^2}w_\pm}{1 - \frac{\lambda_2}{\lambda_1 z_\pm}} \\
    D_{\pm} = \frac{\frac{\lambda_2}{\lambda_1}
    \sqrt{1-\lambda_1^2}z_\pm + \sqrt{1 - \lambda_2^2}w_\pm}{\lambda_1 z_\pm - \lambda_2} + \frac{\sqrt{1 - \lambda_1^2}z_\pm}{\lambda_1} \\
    C_{\pm} = \lambda_1 z_\pm + \sqrt{1 - \lambda_1^2}D_{\pm} \\
    A_{\pm} = z_\pm \\
    B_{\pm} = \frac{F_{\pm}}{z_\pm} 
\end{matrix}
\label{eq:amplitudes}
\end{equation}
where 
\begin{equation}
    z_\pm = e^{i k_\pm L} \qquad {\rm and} \qquad  w_{\pm} = e^{i k_{\pm} 2N} \, .
    \label{eq:def_w_N}
\end{equation} 
Finally, we compute the normalization of the state defined by Eq.~(\ref{eq:psik_2defects_1}). The calculation is long but straightforward, and we find that the squared norm is $2 \left( L + N(|C_\pm|^2-1) \right)$. 

Putting everything together, the normalized eigenstates are
\begin{equation}
    \label{eq:psik_2defects_1}
     \Psi_{k_\pm}(x) = \frac{1}{\sqrt{2 \left( L + N(|C_\pm|^2-1) \right)}} \begin{cases}
         A_\pm e^{i k_\pm x} + B_\pm e^{-i k_\pm (x-1)} \qquad \text{for } x \in [-L/2 +1,0 ] \\
         C_\pm e^{ik_\pm x} + D_\pm e^{-ik_\pm(x-1)} \qquad \text{for } x \in [0,N+1 ] \\
          e^{ikx} +  F_\pm e^{-ik_\pm(x-1)} \qquad\qquad\; \text{for } x \in [N+1,L/2 ] .
     \end{cases}
\end{equation}
Similarly to the one-defect case, since we found $L$ independent solutions and the matrix $h$ is of size $L\times L$, our ansatz exhausts all possible single-particle eigenstates. Having determined the exact eigenstates of the system with periodic boundary conditions, we can now use them to evaluate the propagator $K_t(x,y)$, in direct analogy with the single-defect case studied in Sec.~\ref{sec:singledefect}.

\subsection{Propagator \texorpdfstring{$K_t(x,y)$}{Kt(x,y)}}
Here we present our formulas for the fermion propagator $K_t(x,y)$, defined as in Eq.~(\ref{eq:propagator})
for the double-defect Hamiltonian $\hat{H}_{\lambda_1,\lambda_2}$. We start by giving the final result for the system on the infinite line ($L \rightarrow \infty$).

We introduce the notation $L,C,R$ referring to left, center and right regions respectively:
\begin{equation}
    \begin{array}{rll}
           {\rm left \, }(L) & : \qquad &  x \leq 0 \\
           {\rm center \, }(C) & : \qquad &  0 < x \leq N  \\
           {\rm right \, }(R) & : \qquad &  N < x \, . \end{array}
\end{equation}
The final result for the propagator can be summarized in the following table:
%\newline
\begin{center}
\begin{equation}
\begin{tabular}{ |c|c|c|c| } 
 \hline
 & $y \in L $ & $y \in C $ & $y \in R $ \\
 \hline
 $x \in L $ & $K_{LL}^{\lambda_1\lambda_2}(x,y,t)$ & $[K_{CL}^{\lambda_1\lambda_2}(y,x,-t)]^*$ &  $[K_{RL}^{\lambda_2 \lambda_1}(N+1-x,N+1-y,t)]^*$ \\ 
 \hline
 $x \in C $ & $K_{CL}^{\lambda_1\lambda_2}(x,y,t)$ & $K_{CC}^{\lambda_1\lambda_2}(x,y,t)$ & $-[K_{CL}^{\lambda_2 \lambda_1}(N+1-x,N+1-y,t)]^*$ \\
 \hline
 $x \in R $ & $K_{RL}^{\lambda_1\lambda_2}(x,y,t)$  & $K_{CL}^{\lambda_2\lambda_1}(N+1-y,N+1-x,-t)$ & $[K_{LL}^{\lambda_2 \lambda_1}(N+1-x,N+1-y,t)]^*$ \\ 
 \hline
\end{tabular}
\label{table:prop_2defect}
\end{equation}
\end{center}
where the different functions are defined as follows,
\begin{subequations}
\begin{align}
       K_{LL}^{\lambda_1\lambda_2}&(x,y,t) = i^{x-y} J_{x-y}(t) - \sqrt{1 - \lambda_1^2} i^{x+y-1}J_{x+y-1}(t) \notag \\ 
         & + \lambda_1^2 \sum_{n=0}^{\infty} (-1)^{n+1}\sqrt{1 - \lambda_1^2}^n \sqrt{1 - \lambda_2^2}^{n+1}  i^{x+y-1-2N(n+1)} J_{x+y-1-2N(n+1)}(t)
\end{align}
\begin{align}
       K_{CL}^{\lambda_1\lambda_2}&(x,y,t) = \lambda_1 \sum_{n=0}^{\infty} (-1)^n\sqrt{1 - \lambda_1^2}^n \sqrt{1 - \lambda_2^2}^n i^{x-y+2Nn} J_{x-y+2Nn}(t) \notag \\
         & + \lambda_1 \sum_{n=0}^{\infty} (-1)^{n+1}\sqrt{1 - \lambda_1^2}^n \sqrt{1 - \lambda_2^2}^{n+1} i^{x+y-1-2N(n+1)} J_{x+y-1-2N(n+1)}(t)
\end{align}
\begin{eqnarray}
       K_{RL}^{\lambda_1\lambda_2}(x,y,t) = \lambda_1\lambda_2 \sum_{n=0}^{\infty} (-1)^n\sqrt{1 - \lambda_1^2}^n \sqrt{1 - \lambda_2^2}^n i^{x-y+2Nn} J_{x-y+2Nn}(t),
       \label{eq:K_RL}
\end{eqnarray}
\begin{comment}
\begin{eqnarray}
       K_{CC}^{\lambda_1\lambda_2}(x,y,t) = i^{x-y}J_{x-y}(t) + \sum_{n =1}^\infty \sqrt{1 - \lambda_1^2}^n \sqrt{1 - \lambda_2^2}^n (-1)^n  \nonumber \\
       \left[ i^{x-y+2Nn} J_{x-y+2Nn}(t) + i^{x-y-2Nn} J_{x-y-2Nn}(t) \right] + \nonumber \\
       \sum_{n =0}^\infty \sqrt{1 - \lambda_1^2}^{n+1} \sqrt{1 - \lambda_2^2}^n (-1)^n i^{x+y-1+2Nn} J_{x+y-1+2Nn}(t) +  \nonumber \\
       \sum_{n =1}^\infty \sqrt{1 - \lambda_1^2}^n \sqrt{1 - \lambda_2^2}^{n+1} (-1)^{n+1} i^{x+y-1-2Nn} J_{x+y-1-2Nn}(t).
\end{eqnarray}
\end{comment}
\begin{eqnarray}
       K_{CC}^{\lambda_1\lambda_2}(x,y,t) = \sum_{n=0}^\infty \sqrt{1 - \lambda_1^2}^n \sqrt{1 - \lambda_2^2}^n (-1)^n {\large[} i^{x-y+2Nn}J_{x-y+2Nn}(t) \nonumber \\ + i^{x-y-2Nn}J_{x-y-2Nn}(t) + \sqrt{1 - \lambda_1^2} i^{x+y-1+2Nn} J_{x+y-1+2Nn}(t)\nonumber \\ -\sqrt{1 - \lambda_2^2} i^{x+y-1+2N(n+1)} J_{x+y-1+2N(n+1)}(t){\large]}  - i^{x-y}J_{x-y}(t).
\end{eqnarray}
\label{eq:prop_2defect}
\end{subequations}

Let us sketch the derivation of these formulas. For concreteness we focus on the case of $K_{RL}^{\lambda_1 \lambda_2}(x,y,t)$ as an example. The other terms can be obtained following the same steps.

\begin{figure}[t]
    \centering
    \includegraphics[width = \textwidth]{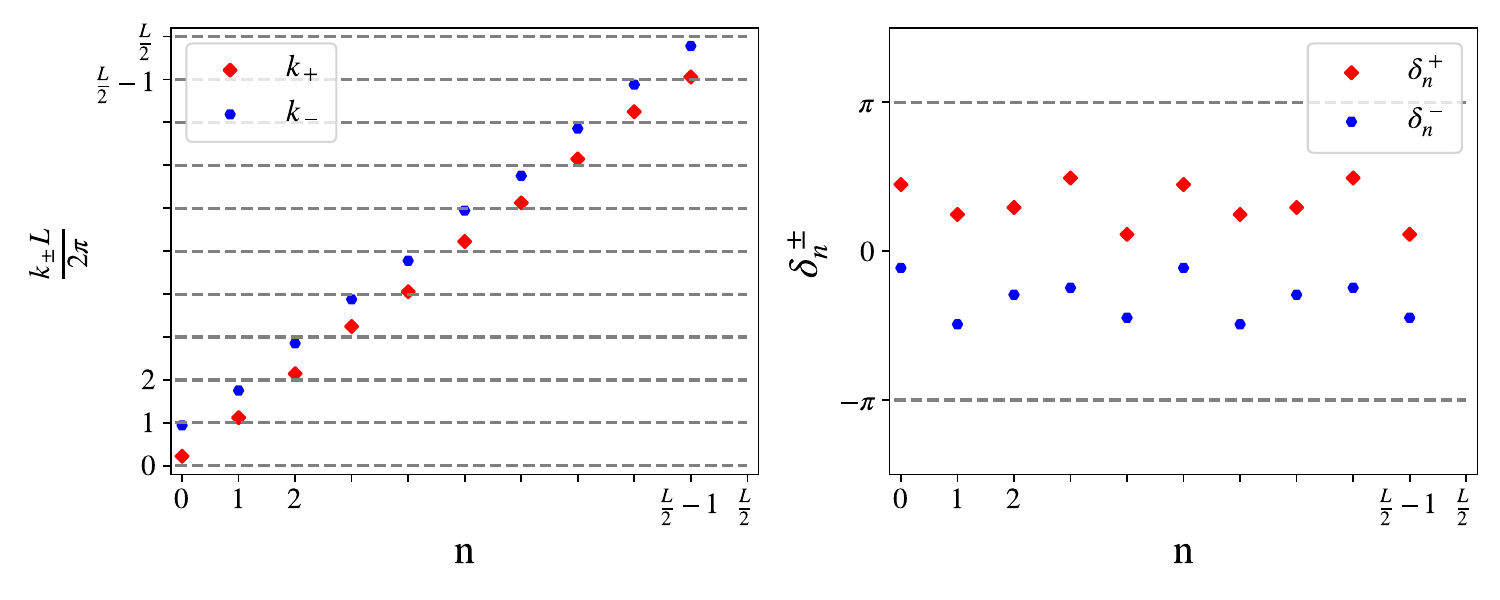}
    \caption{Left: The two solutions $k_+$ (red) and $k_-$ (blue) of the quantisation condition \eqref{eq:quant_cond_2def_4}. Horizontal dashed lines show the values in case of a clean system (without the defects). Right:
    The parameters $\delta_n^+$ (red) and $\delta_n^-$ (blue)  introduced in Eq. \eqref{eq:2defect_density_hydrolimit}. The values for the quasimomenta $k_\pm$ were obtained by exact diagonalization for a system of size $L = 20$, for defect parameters $\lambda_1 = 0.77$ and $\lambda_2 = 0.63$.
    }
    \label{fig:quasimomenta_2defect}
\end{figure}

Starting from  the exact single-particle eigenstates \eqref{eq:psik_2defects_1}, we get
\begin{multline}
 K_{RL}^{\lambda_1 \lambda_2}(x,y,t) = \sum_{\alpha = +,-} \sum_{k_\alpha} e^{i \varepsilon(k_\alpha) t} \Psi_{k_\alpha}(x)\Psi^*_{k_\alpha}(y) \\
       = \sum_{\alpha = +,-} \sum_{k_\alpha} e^{i \varepsilon(k_\alpha) t} \frac{ \frac{1}{z_\alpha} e^{ik_\alpha(x-y)} + z_\alpha e^{ik_\alpha(y-x)} + \frac{F_\alpha}{z_\alpha} e^{-ik_\alpha (x+y-1)} + z_\alpha F^*_\alpha e^{ik_\alpha (x+y-1)} }{2 \left( L + N(|C_\alpha|^2-1) \right)} . 
       \label{eq:example}
\end{multline}
Here we have used the fact that the amplitude $F_\alpha$, defined in Eq.~\eqref{eq:amplitudes}, is a pure phase, i.e.\ $|F_\alpha|^2 = 1$. The sum runs over the two sets of solutions to the quantization condition (\ref{eq:quant_cond_2def_4}). In Fig.~\ref{fig:quasimomenta_2defect} we show a typical configuration of such solutions, for some generic values of the parameters $\lambda_1, \lambda_2$. The solutions are always of the form
\begin{equation}
    k_\pm = \frac{2\pi m}{L} + \frac{\delta_{m,\pm}}{L} \qquad {\rm where} \qquad |\delta_{m,\pm}| < \pi \, \qquad {\rm for} \quad m=0,\dots, \frac{L}{2}-1 .
    \label{eq:quant_cond_2def_2}
\end{equation}
Using this expression we can evaluate the $L \to \infty$ limit of Eq.~(\ref{eq:example}). Going term by term; let us first focus on the second term,
\begin{eqnarray}
    \label{eq:intermediate1}
       \sum_{\alpha = +,-} \sum_{k_\alpha} e^{i \varepsilon(k_\alpha) t} \frac{ z_\alpha e^{ik_\alpha(y-x)} }{2 \left( L + N(|C_\alpha|^2-1) \right)} .
\end{eqnarray}
Taking the thermodynamic limit $L \rightarrow \infty$ while keeping $N, x,y$ and $t$ constant, the correction term $ \propto \delta_{m,\pm}$ in the quasimomenta $k_\pm$ is negligible everywhere except in $z_\pm = e^{i k_\pm L}$. Then one can perform the summation over $\alpha = +,-$, noting that
\be
z_+ + z_-   \underset{L \rightarrow \infty}{ \longrightarrow}   \frac{2 \lambda_1 \lambda_2}{1 + \sqrt{1 - \lambda_1^2}\sqrt{1 - \lambda_2^2}[w_\pm]^*} 
  = 2 \lambda_1 \lambda_2 \sum_{n = 0}^\infty (-1)^n \sqrt{1 - \lambda_1^2}^n \sqrt{1 - \lambda_2^2}^n \left[\frac{1}{w_\pm}\right]^n  \, ,
\ee
where we obtain the last form by recognising that the large $L$ limit takes the form of a sum of a geometric series. The expression (\ref{eq:intermediate1}) becomes, when $L \rightarrow \infty$,

\begin{eqnarray}
    \label{eq:intermediate2}
      \sum_{m=0}^{L/2-1} e^{i \varepsilon( \frac{2\pi m}{L} ) t} \frac{ (z_+ + z_-) e^{i \frac{2\pi m}{L} (y-x)} }{ 2L }  \,\underset{L \rightarrow \infty}{ \longrightarrow} \,& \lambda_1 \lambda_2 \sum_{n=0}^\infty  (-1)^n \sqrt{1 - \lambda_1^2}^n \sqrt{1 - \lambda_2^2}^n \times \\ \nonumber
      \times \int_0^{\pi} {\rm d}k \frac{1}{2 \pi} e^{i \varepsilon(k) t } e^{i k (y-x)} e^{-ik2Nn}.
\end{eqnarray}
Here we used \eqref{eq:quant_cond_2def_2} to write the quasimomenta as
\begin{equation}
    k_\pm \approx \frac{2 \pi m}{L} + \mathcal{O}(L^{-1}) \text{,} \quad \text{with} \quad m = 0,1,...,L/2-1.
\end{equation}
where the second term $\mathcal{O}(L^{-1})$ only gives a non-negligible contribution in the exponent if it is $\propto k_\pm L$, in which case it can be evaluated by direct substitution of the formula for $z_\pm$ in Eq \eqref{eq:def_w_N}. Furthermore, replacing the sum in the thermodynamic limit as $\sum_m \rightarrow \int \frac{L}{2 \pi} {\rm d}k $, and directly substituting the formula for $w_\pm$ from Eq. \eqref{eq:def_w_N}, then transforming the integrand as $q = -k$, allows us to evaluate the integral in terms of Bessel functions:
\begin{eqnarray}
       \int_{-\pi}^{\pi} \frac{{\rm d}k}{2 \pi} e^{i k (x-y + 2Nn) - i\cos(k)t} = J_{x - y + 2Nn}(t) i^{x-y+2Nn}.
\end{eqnarray}
It is straightforward to see following these steps that the evaluation of first term in \eqref{eq:example} follows the same calculation and results in in the same term. 

Continuing the evaluatation of the full propagator we turn to the third and forth terms in the expression \eqref{eq:example}. Following the same ideas as for the other terms we can perform the summation over $\alpha = +,-$. The result gives
\begin{equation}
    \quad \sum_\alpha \frac{F_\alpha}{z_\alpha} = 0.
\end{equation}
This means that there are no contributions to the propagator coming from the third and fourth terms in \eqref{eq:example}.
Putting the terms together results in the formula \eqref{eq:K_RL} for the propagator.
This calculation gives a clear recipe for obtaining the rest of the propagators presented above: using the formula \eqref{eq:propagator_def} one obtains the propagator in terms of the amplitudes \eqref{eq:amplitudes}, where one can evaluate the summation over the two different species by pairing the quasimomenta that correspond to the same quantum number $m$. The resulting amplitudes can always be written as geometric series. Taking the thermodynamic limit then as discussed above the integral over the quasimomenta $k$ can be evaluated resulting in the propagators as sums of Bessel functions.

\subsection{Semiclassical picture}
\label{sec:semiclassical_2def}
As shown in Table \eqref{table:prop_2defect} and Eqs. \eqref{eq:prop_2defect}, the propagator of a particle $K^{\lambda_1 \lambda_2}(x,y,t)$ is given by a sum of Bessel functions of the first kind. Deriving a semiclassical picture for the particle dynamics can be done by carefully examining the terms present in the propagators. 
\begin{figure}
    \centering
    \includegraphics[width = 0.5\textwidth]{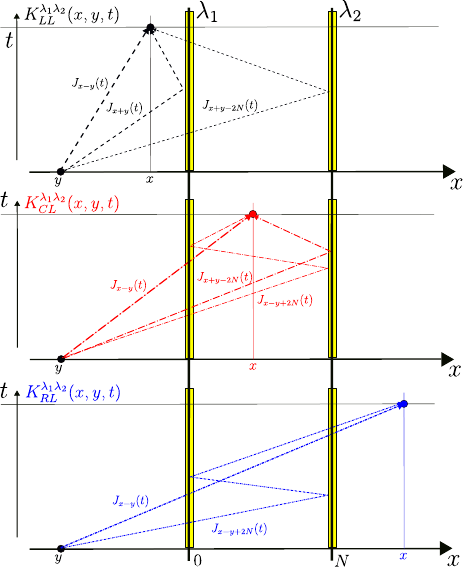}
    \caption{Schematic drawing of the semiclassical picture of the propagators for different positions of x ($x \in L,C,R $ respectively) and $y \in L$, with time being the vertical axis and coordinate the horizontal, with defect strengths of $\lambda_1, \lambda_2 \in [0,1]$ at $x = 0,N$ repectively.}  
    \label{fig:2defect_classical_paths}
\end{figure}
 
Taking for example the case of $x,y \in [-\frac{L}{2} + 1, 0]$, the propagator is given by $K_{LL}^{\lambda_1\lambda_2}(x,y,t)$ in Eq. \eqref{eq:prop_2defect}, in which we see three different terms describing different classical paths. The first term being $\propto J_{x-y}(t)$ represents a propagation from point $y$ to $x$ without being affected by either of the defects, with an additional phase given by the length of the propagation. %Note that the length here takes a sign as both $x,y \leq 0$. 
The second term is $\propto J_{x+y-1}(t)$, with $x+y-1$ being the length of the classical path of the particle from $y$ to $x$ with a reflection at position $1$. This term has the same phase coming from the classical path as well as an additional sign coming from the fact that the reflection causes the particle to go in the opposite direction. The amplitude $\sqrt{1 - \lambda_1^2}$ can be understood classically as the probability that the particle got reflected at the defect. 
The last term is given by an infinite sum of terms of the form 
\begin{equation}
    \lambda_1^2  (-1)^{n+1}\sqrt{1 - \lambda_1^2}^n \sqrt{1 - \lambda_2^2}^{n+1} i^{x+y-1-2N(n+1)} J_{x+y-1-2N(n+1)}(t).
\end{equation}
It describes a classical path of a particle from position $y$ reaching the first defect (with parameter $\lambda_1$) scattering on it, then getting transmitted, giving a factor $\lambda_1$ in the amplitude. After that the particle bounces in the central region $n$ times giving the factors  $\sqrt{1 - \lambda_1^2}^n \sqrt{1 - \lambda_2^2}^{n}$ in the amplitude from the multiple reflections, finally getting reflected on the second defect and transmitted on the first one, hence the remaining factor $\lambda_1 \sqrt{1 - \lambda_2^2}$. The phases, like in the previous cases are given by the (signed) distance travelled by the particle (giving the factor $i^{x+y-1-2N(n+1)}$) as well as the number of times the particle changed to the opposite direction (giving the factor $(-1)^{n+1})$). The classical paths emerging from this semiclassical interpretation are shown on Fig. \ref{fig:2defect_classical_paths} for the examples of $K_{LL}^{\lambda_1\lambda_2}(x,y,t)$, $K_{CL}^{\lambda_1\lambda_2}(x,y,t)$ and $K_{RL}^{\lambda_1\lambda_2}(x,y,t)$ respectively.

%In the following section we show how the propagators can be used to obtain corrections to the naive hydrodynamic picture for a density profile of fermion being released from a domain wall initial state and let freely expand.

\section{Density profiles with two defects after a domain-wall release}
\label{sec:density}
At $t = 0$, the system is prepared in the product state~\eqref{eq:DW_initstate}. This corresponds to filling the left side of the system up to and including $x = 0$, where the first defect, characterized by the parameter $\lambda_1$, is located. At $t > 0$ the system evolves with the Hamiltonian \eqref{eq:hoppingH2} with hopping \eqref{eq:2defects} containing the two defects. 
The density profile is given by
\begin{equation}
    \rho(x,t) = \sum_{y\leq 0} K_t(x,y) K^*_t(x,y).
\end{equation}
Using the formulas derived for the propagator in Table \eqref{table:prop_2defect} and Eqs. \eqref{eq:prop_2defect}, the density profile for the free expansion is given by:
\begin{subequations} \newline for $x \leq 0$:
\begin{align}
    \rho(x,t) &= \sum_{y\leq 0} J_{x-y}(t)^2 + (1 - \lambda_1^2) J_{x+y-1}(t)^2 \notag \\
    &\quad + 2\lambda_1^2 \sum_{n=0}^\infty (-1)^{N(n+1)+n} \sqrt{1-\lambda_1^2}^{n+1} 
    \sqrt{1-\lambda_2^2}^{n+1} 
  %  \notag \\
  %  &\qquad \times
  J_{x+y-1}(t)J_{x+y-1-2N(n+1)}(t)
    \notag \\
    &\quad + \lambda_1^4\sum_{n,m=0}^\infty (-1)^{(N+1)(n+m)} \sqrt{1-\lambda_1^2}^{n+m} 
    \sqrt{1-\lambda_2^2}^{n+m+2}
  %  \notag \\  &\qquad \times
  J_{x+y-1-2N(n+1)}(t)J_{x+y-1-2N(m+1)}(t),
\end{align}
for $1 \leq x \leq N$:
\begin{align}
    \rho(x,t) &= \lambda_1^2 \sum_{y \leq 0}\sum_{n,m=0}^\infty (-1)^{(N+1)(n+k)} \sqrt{1 - \lambda_1^2}^{n+k} \sqrt{1 - \lambda_2^2}^{n+k}{\large(} J_{x-y+2Nn}(t)J_{x-y+2Nm}(t)  \notag \\
    &\quad +  \sqrt{1 - \lambda_2^2}^{2} J_{x+y-1-2N(n+1)}(t)J_{x+y-1-2N(m+1)}(t) {\large)}
\end{align}
and for $N < x$:
\begin{align}
    \rho(x,t) &= \lambda_1^2 \lambda_2^2 \sum_{y \leq 0} \sum_{n,m=0}^{\infty} 
    \sqrt{1 - \lambda_1^2}^{n+m} \sqrt{1 - \lambda_2^2}^{n+m} (-1)^{(N+1)(n+m)} %\notag \\ &\quad \times 
    J_{x-y+2Nn}(t) J_{x-y+2Nm}(t).
\end{align}
\label{eq:2defect_density}
\end{subequations}

\begin{figure}
    \centering
    \includegraphics[width = \textwidth]{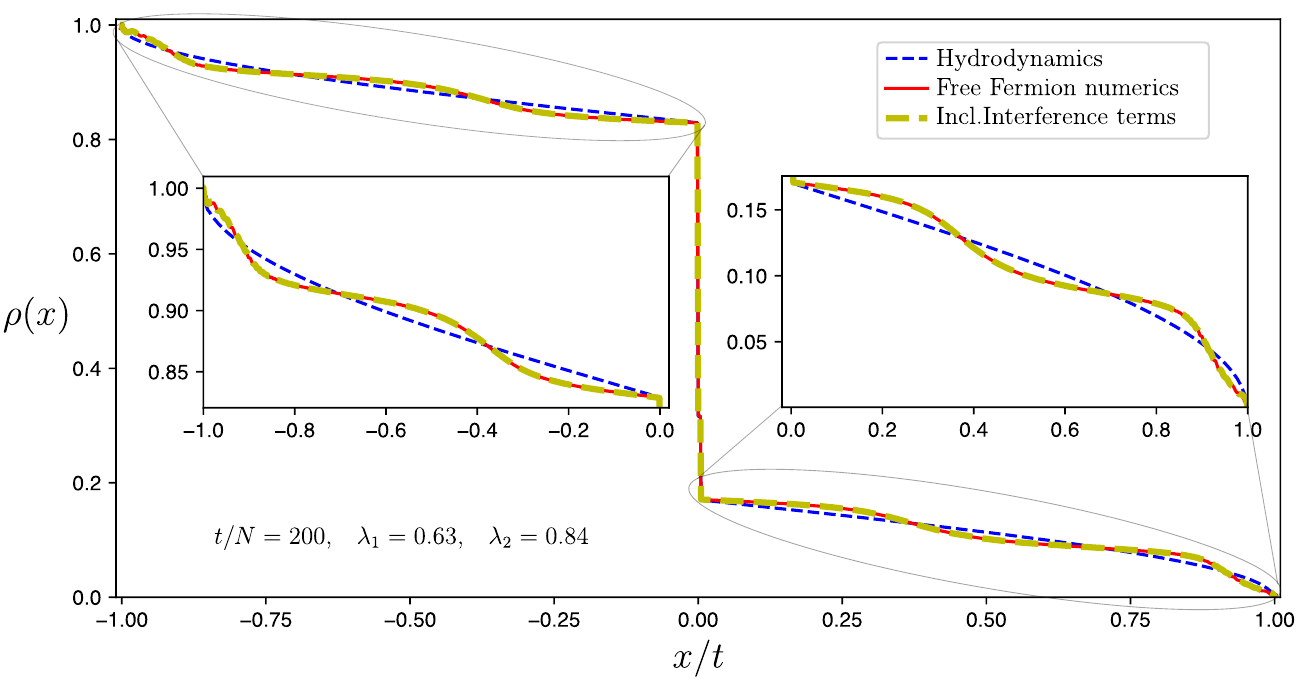}
    \caption{Density profile of the free expansion from the domain-wall initial state in Eq.~\eqref{eq:DW_initstate}, at a time $t \gg N$ and for defect strengths $\lambda_1 = 0.63$ and $\lambda_2 = 0.84$. We compare the analytical expression for the density (green dashed line) given in Eq.~\eqref{eq:2defect_density_hydrolimit} with exact free-fermion numerics (red solid line) and  a 'naive' hydrodynamic prediction (blue dashed line)  that is obtained by treating the setup as two independent scatterers, each described by the quasi-particle picture discussed in Section \ref{section:Quasi_part_prop} (this description doesn't account for the 'interference terms' in \eqref{eq:2defect_density_hydrolimit}, keeping only terms diagonal in the sums over multiple reflections) . The insets highlight the regions to the left and right of the defects, where deviations from the hydrodynamic description are most pronounced.}
    \label{fig:density_eulerlimit}
\end{figure}

Taking the Euler scaling limit allows us to evaluate the diagonal terms (i.e $n = m$) in Eqs. \eqref{eq:2defect_density} by approximating $\sum_{y<0} \approx \int_{-\infty}^0 {\rm d}y $. Making use again of the saddle point approximation of the Bessel functions given in \eqref{eq:Bessel_straddle} results in the density
\begin{subequations}
\begin{align}
    \rho(x,t) &= 1 - \frac{\lambda_1^2}{\pi} \arccos\left(\frac{-x}{t}\right) 
          {\rm \quad  } + \frac{\lambda_1^4}{\pi} \sum_{n = 0}^{\lfloor \frac{t + x}{2N} \rfloor} (1 - \lambda_1^2)^{n}(1 - \lambda_2^2)^{n+1}   \arccos\left(\frac{2nN-x}{t}\right) \notag \\
    &\quad + 2\lambda_1^2 \sum_{n=0}^\infty (-1)^{N(n+1)+n} \sqrt{1-\lambda_1^2}^{n+1} 
    \sqrt{1-\lambda_2^2}^{n+1} 
    %\notag \\ &\qquad \times 
    J_{x+y-1}(t)J_{x+y-1-2N(n+1)}(t)
    \notag \\
    &\quad + 2\lambda_1^4\sum_{n<m}^\infty (-1)^{(N+1)(n+m)} \sqrt{1-\lambda_1^2}^{n+m} 
    \sqrt{1-\lambda_2^2}^{n+m+2}
    \notag \\ &\qquad \times 
    J_{x+y-1-2N(n+1)}(t)J_{x+y-1-2N(m+1)}(t),
\end{align}
for $x \leq 0$, 
\begin{align}
    \rho(x,t) &= \frac{\lambda_1^2}{\pi} \sum_{n=0}^{\lfloor  \frac{t - x}{2N} \rfloor} 
    (1 - \lambda_1^2)^{n} (1 - \lambda_2^2)^{n} 
    \arccos\left(\frac{x+2nN}{t}\right) \notag \\
    &\quad + \frac{\lambda_1^2}{\pi} \sum_{n=0}^{\lfloor  \frac{t + x}{2N} \rfloor} 
    (1 - \lambda_1^2)^{n} (1 - \lambda_2^2)^{n+1}  \arccos\left(\frac{2(n+1)N-x}{t}\right) \notag \\  
    &\quad + 2 \lambda_1^2 \sum_{y \leq 0} \sum_{n < m}(-1)^{(N+1)(n+m)}
    \sqrt{1 - \lambda_1^2}^{n+m} \sqrt{1 - \lambda_2^2}^{n+m} 
     \notag \\  
    &\qquad \times \left[ J_{x-y+2Nn}(t) J_{x-y+2Nm}(t) + (1 - \lambda_2^2) 
    J_{x+y-1-2N(n+1)}(t) J_{x+y-1-2N(m+1)}(t) \right],
\end{align}
for $1 \leq x \leq N$ and 
\begin{align}
    \rho(x,t) &= \frac{\lambda_1^2\lambda_2^2}{\pi} 
    \sum_{n=0}^{\lfloor  \frac{t - x}{2N} \rfloor}  
    (1 - \lambda_1^2)^{n} (1 - \lambda_2^2)^{n}    
    \arccos\left(\frac{x+2nN}{t}\right) \notag \\  
    &\quad + 2 \lambda_1^2 \lambda_2^2 
    \sum_{y \leq 0} \sum_{n < m}
    (-1)^{(N+1)(n+m)} \sqrt{1 - \lambda_1^2}^{n+m} 
    \sqrt{1 - \lambda_2^2}^{n+m} 
   % \notag \\ &\qquad \times 
    J_{x-y+2Nn}(t) J_{x-y+2Nm}(t),
\end{align}
for $N < x$.
\label{eq:2defect_density_hydrolimit}
\end{subequations}
We emphasize that, in the expressions given in Eq.~\eqref{eq:2defect_density_hydrolimit}, Euler-scale asymptotics can be obtained only for the diagonal contributions, while the terms involving $n \neq m$ do not admit a simple Euler-scale limit. We compare these analytical results with exact free-fermion numerical calculations in Fig.~\ref{fig:density_eulerlimit}. It is also important to note that retaining only the diagonal terms in $n$ in the propagator~\eqref{eq:prop_2defect} reproduces the result of the naive hydrodynamic picture. This shows that the off-diagonal terms encode the corrections arising from quantum-interference effects, which can be understood within the semiclassical interpretation.

Figure~\ref{fig:density_eulerlimit} summarizes the main findings of this work. The hydrodynamic prediction (blue dashed curve) fails to reproduce the exact numerical density profile, as it does not capture the interference effects generated by the two defects. Once these interference contributions are properly taken into account, the agreement with the exact numerics becomes essentially perfect.

%\subsection{The large-time limit}
\subsection{ The correct hydrodynamic limit}
We have shown that the dynamics cannot be adequately described by a hydrodynamic picture, as is possible in the case of a single defect or in the absence of defects. This is because, at finite times, no closed evolution equation exists for the corresponding Wigner function.
For large distances and times, however, such that
\begin{equation}
    N/t \rightarrow 0, \quad {\rm and } \quad L/t, \quad x/t \rightarrow {\rm fixed},
\end{equation}
the dynamics become classical again and one can write down an evolution equation for the Wigner function in the form of Eq. \eqref{eq:domain_wall_occupation_func}:
\be
\nu(x,k,t) = \begin{cases}
    1  & \quad \text{ if } \; x \leq \min[\sin(k)t,0] \\
    1 - T_{\lambda_1,\lambda_2}(k) &  \quad \text{ if } \;  \min[\sin(k)t,0] < x \leq 0 \\
    T_{\lambda_1,\lambda_2}(k)&  \quad \text{ if } \;  0 < x \leq \max[\sin(k)t,0] \\
    0 & \quad \text{ otherwise},
\end{cases}
\label{eq:occupation_func_infT}
\ee
where $T_{\lambda_1,\lambda_2}(k)$ is a quasi-momentum dependent transmission amplitude that is a function of the defect parameters $\lambda_1$ and $\lambda_2$.
Eq.~\eqref{eq:occupation_func_infT} shows that, after a sufficiently long time $t \gg N$, the two defects can be effectively treated as a single composite defect. Physically this behavior can be understood as a consequence of the fact that for large time we have many multiple reflections between the two defects. In this regime, interference effects are progressively washed out by the large number of contributing paths, resulting in an \emph{effectively classical} behavior of the system.

\begin{figure}
    \centering
    \includegraphics[width = 0.8\textwidth]{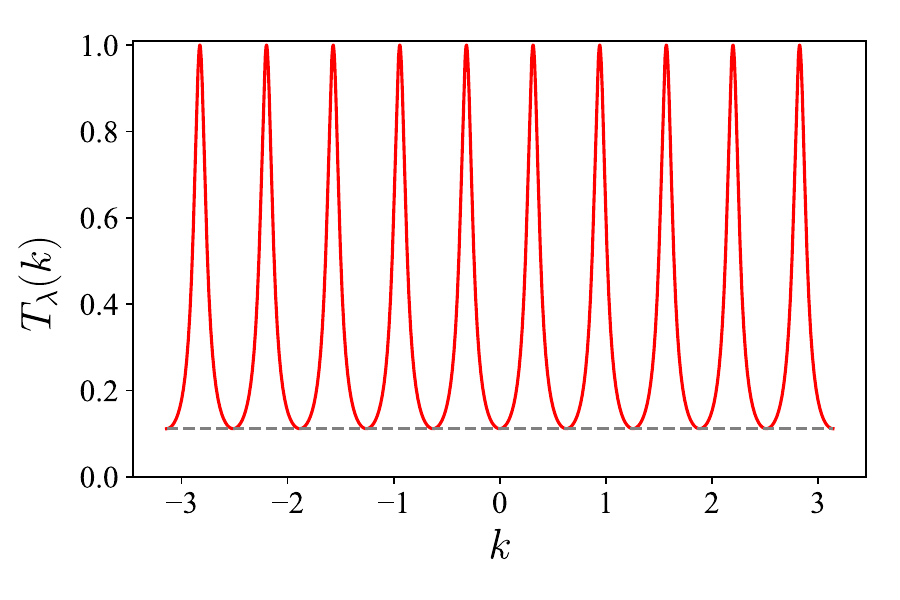}
    \caption{The transmission amplitude $T_\lambda(k)$ as a function of quasimomentum $k$ given in Eq. \eqref{eq:T_double_defect} for $\lambda^2 = 1/2$, $N = 5$ with $T_{min} = \frac{\lambda^4}{(2 - \lambda^2)^2}$ plotted with grey dashed line (at every $kN = n\pi$ for $n \in \mathbb{Z} $) and $T_{max} = 1$ (for every $kN = \frac{\pi}{2} + n\pi$ for $n \in \mathbb{Z}$ )}
    \label{fig:my_label}
\end{figure}

\begin{figure}[t]
    \centering
    \includegraphics[width = 0.9 \textwidth]{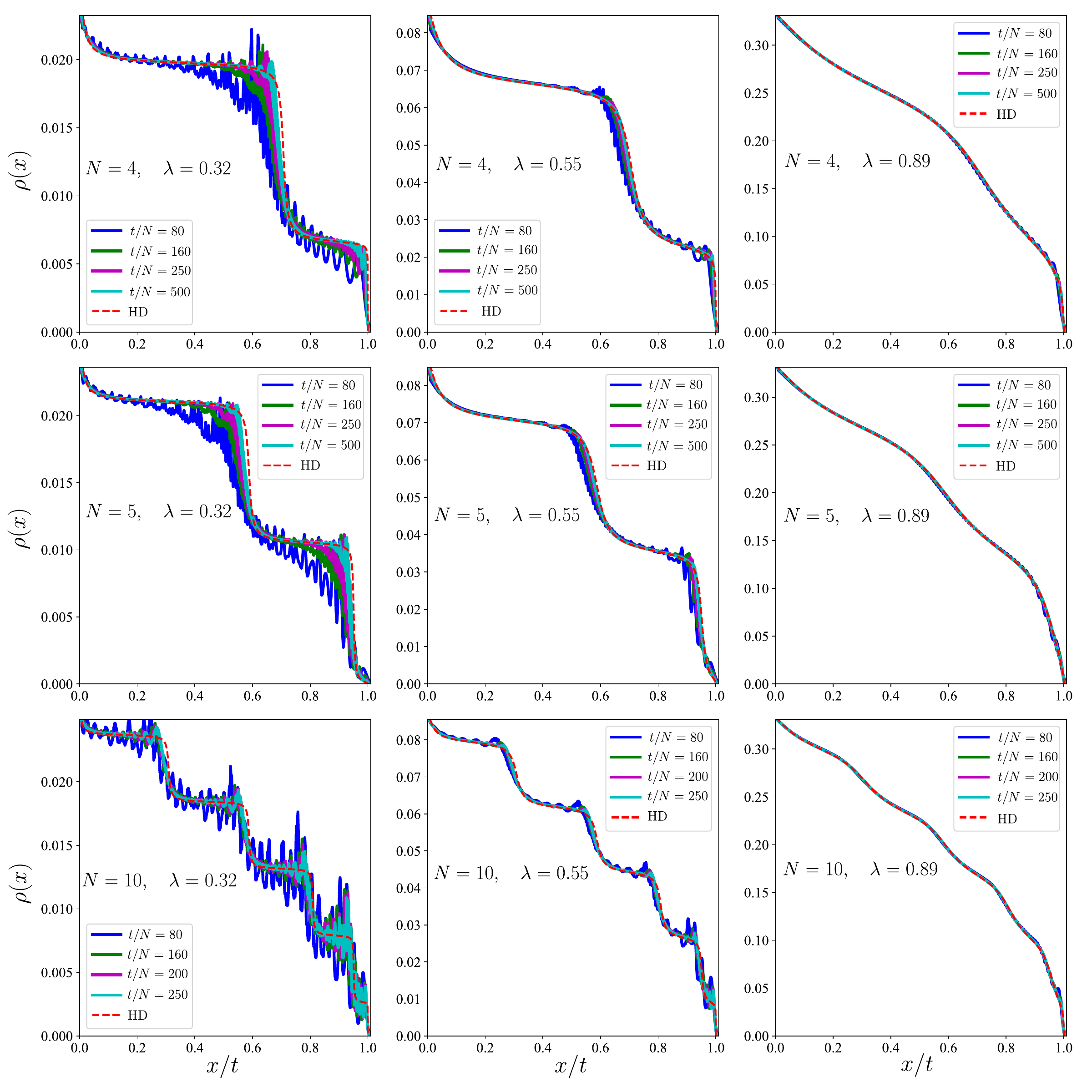}
    \caption{Convergence of the density profile to the infinite-time limit on the right side of the composite defect ($N < x$). Results are shown for defect separations $N = 4, 5, 10$ (from top to bottom) and for coupling strengths $\lambda^2 = 0.1, 0.3, 0.8$ (from left to right). Solid lines correspond to numerical data at different times, with the horizontal axis rescaled by the time $t$. These are compared with the hydrodynamic prediction in Eq.~\eqref{eq:density_asymptotic}, shown as red dashed lines.}
    \label{fig:infT_transmission}
\end{figure}

In the remaining part of this section we assume $\lambda_1 = \lambda_2 = \lambda$ for simplicity. The results generalize straightforwardly to the case of different defect parameters.
The transmission amplitude $T_{\lambda,\lambda}(k)$ (just $T_{\lambda}(k)$ in the following) can be obtained by solving the scattering problem on the composite defect as follows: An incident plane wave with quasimomentum $k > 0$ coming from the left $x \ll 0$ scatters on the composite defect, which results in a reflected and a transmitted component, thus, in the limit of infinite times, the ansatz for the wave function is
\begin{equation}
    \Phi(x) = \frac{1}{\sqrt{L}}\begin{cases}
        A e^{iqx} + Be^{-iq (x-1)}, & \qquad \text{for } \quad x \leq 0
        \\
        e^{iqx}, & \qquad \text{for } \quad \frac{N}{2} \leq x.
    \end{cases} 
\end{equation}
The equations for amplitudes $A$ and $B$ can be obtained by, similarly to Eqs. \eqref{eq:matrices_M}, writing the conditions on $\Phi(x)$ arising from the two defects
\begin{equation}
    \begin{pmatrix}
        1 \\
        0
    \end{pmatrix} = \frac{1}{\lambda^2} \begin{bmatrix}
        1 & -\sqrt{1 - \lambda^2}e^{-i2Nk} \\
        -\sqrt{1 - \lambda^2}e^{i2Nk}  & 1
    \end{bmatrix} \begin{bmatrix}
        1 & -\sqrt{1 - \lambda^2} \\
        -\sqrt{1 - \lambda^2} & 1
    \end{bmatrix}\begin{pmatrix}
      A \\
      B
    \end{pmatrix}.
\end{equation}
This system of two equations results in
\begin{equation}
    A = \frac{1}{\lambda^2}\left( 2 \cos{Nq} - \lambda^2 e^{iNq} \right) \qquad \text{and} \quad B = \frac{e^{iq}}{\lambda^2} \sqrt{1 - \lambda^2} 2 \cos{Nq},
\end{equation}
giving the transmission amplitude
\begin{equation}
    T_\lambda(k) = \frac{1}{|A|^2} =  \frac{1}{\frac{4(1 - \lambda^2)}{\lambda^4}\cos^2{Nk} + 1},
    \label{eq:T_double_defect}
\end{equation}
which is now dependent on quasi-momentum $k$. This composite transmission amplitude is plotted as function of the momentum $k$ in Fig. \ref{fig:my_label} for $\lambda=2^{-1/2}$.
The density profile $\rho(x,t)$ at time $t$ can be now calculated from the Wigner function straightforwardly, cf. Eq. \eqref{eq:density} 
which results in 
\begin{equation}
    \rho(x,t) = \begin{cases}
        0, & \qquad \text{if } \quad t < x \\ 
        \frac{1}{2\pi}\int_{\arcsin\left(\frac{x}{t}\right)} ^{\pi - \arcsin\left(\frac{x}{t}\right)} {\rm d}k T(k), & \qquad \text{if }  0 \leq x  \leq t \\
        1 + \frac{1}{2\pi}\int_{-\pi - \arcsin\left(\frac{x}{t}\right)} ^{ \arcsin\left(\frac{x}{t}\right)} {\rm d}k T(k), & \qquad \text{if }  -t\leq x < 0 \\
        1, & \qquad \text{if } \quad x < -t.
    \end{cases}
    \label{eq:density_asymptotic}
\end{equation}
We compare this prediction with exact free-fermion numerics in Fig.~\ref{fig:infT_transmission} for several large values of $t/N$. The comparison shows a clear convergence toward our analytical result as the time increases.

\section{Conclusion}
\label{sec:concl}

In this work, we have investigated the non-equilibrium dynamics of hard-core bosons released from a domain-wall initial state in the presence of two conformal defects. We have shown that, in this setting, the Euler-scale dynamics differ markedly from the predictions of the generalized hydrodynamics description. This breakdown of hydrodynamics originates from quantum-interference effects generated by repeated scattering processes between the defects, which persist even at large scales and finite times.
These interference phenomena can be understood qualitatively, and captured quantitatively, through a detailed analysis of the exact fermionic propagators. By deriving these propagators explicitly, we demonstrated that a meaningful semiclassical description can still be recovered, provided that the scattering picture accounts for coherent multiple reflections. This extends the approach previously developed for systems containing a single conformal defect~\cite{Capizzi_2023} and suggests that the method generalizes to higher numbers of defects and to more general (i.e. non-conformal) defects.

We derived closed analytic expressions for the boson density profile during free expansion, both at short times and in the Euler (large-time) limit. These results were benchmarked against exact free-fermion numerics, with which we found perfect agreement. Our analysis thus provides a comprehensive and fully controlled description of the dynamics, firmly establishing the role of interference effects beyond conventional hydrodynamics. The present work can be viewed as a natural extension of Ref.~\cite{Capizzi_2023}, where the time evolution of the entanglement entropy was studied in the context of domain-wall melting across a single conformal defect.

Beyond these concrete results, our findings raise several intriguing open questions. In particular, it would be highly interesting to understand how interference corrections can be incorporated in a systematic way into the hydrodynamic framework and whether a generalized formulation of GHD can be developed to account for such effects. Another important direction concerns the extension of this approach to more complex observables, such as bipartite entanglement and other measures of quantum correlations, where interference effects are expected to play an even more pronounced role.

\section*{Acknowledgments}
PC acknowledges support
by the ERC-AdG grant MOSE No. 101199196. AT and JD acknowledge support from ANR-PRME Uniopen (project ANR-22-CE30-0004-01) and from M\'esocentre EXPLOR of  Universit\'e de Lorraine (project 2024CPMXX3457).

\bibliography{momentum_dist}

\end{document}